\documentstyle[aps,prl,floats,psfig]{revtex}
\tighten
\draft
\begin{document}
\twocolumn[\hsize\textwidth\columnwidth\hsize\csname
@twocolumnfalse\endcsname

\title{Collective T=0 pairing in N=Z nuclei? \\ Pairing vibrations around $^{56}$Ni revisited.}

\author{A.O.~Macchiavelli, P.~Fallon, R.M.~Clark, M.~Cromaz,
  M.A.~Deleplanque, R.M.~Diamond,
 G.J.~Lane, I.Y.~Lee, F.S.~Stephens, C.E.~Svensson, K.~Vetter and D.~Ward}
\address{ Nuclear Science Division, Lawrence Berkeley National Laboratory,
Berkeley, CA 94720\\}
\date{\today}
\maketitle

\begin{abstract}

We present a new analysis of the pairing vibrations around
$^{56}$Ni, with emphasis on odd-odd nuclei. 
This analysis of the experimental excitation energies is based on the subtraction of 
average properties that include 
the full symmetry energy together with the  volume, surface, and Coulomb terms. 
The results clearly indicate a  collective 
behavior of the isovector pairing 
vibrations and do not support any appreciable collectivity in the isoscalar channel.
\end{abstract}
\vspace{0.5cm}]


\narrowtext 

Driven by recent advances in experimental techniques and the new possibilities that will become
available with radioactive beams, we are witnessing a revival of nuclear structure studies 
along the $N=Z$ line. Of particular interest in these nuclei is the role played by the
 isoscalar $(T=0)$ and isovector $(T=1)$ pairing correlations.
Given the charge independence of the nuclear force, we expect to observe the effects of
the standard $T=1$ pairing on an equal footing between the $T_z=0$ ($np$) and $|T_z|=1$ ($
nn$ and $pp$) components. In addition, we have the unique possibility of studying the 
formation of a condensate of $T=0$ $np$ Cooper pairs.

To address some of these interesting questions we analyzed, in a recent work\cite{aom}, the experimental binding energies of nuclei along the
$N=Z$ line with emphasis on the lowest $T=0$ and $T=1$ states in 
self-conjugate odd-odd nuclei. Our analysis indicates that the odd-odd $T=1$ states are as bound as the ground-state in
their even-even neighbors, and this can be interpreted as a consequence of full isovector pairing\cite{aom,vogel}. The odd-odd $T=0$ states were found to be less bound than the even-even neighbors by an energy $2\Delta_{T=1} \approx 2(12$MeV$/\sqrt A)$
which amounts to the blocking of the $T=1$ correlations.  This result  not only
suggests that the $T=0$ states
 behave like any other odd-odd nucleus ground state ($T_z \ne 0$) in the nuclear chart, but also 
leaves no room to support 
the existence of an isoscalar pair condensate\cite{aom}.

Near closed shells, 
the strength of the pairing force relative to the single-particle level-spacing
is expected to be less than the critical value needed to obtain a 
superconducting solution,  and the pairing field then gives rise to a collective
phonon\cite{bes1}.  Transfer reactions provide a characteristic fingerprint for the 
collectivity of this elementary mode of excitation which has been studied in detail
around $^{208}$Pb\cite{bes1,bm1}.

In 1969, A.Bohr \cite{bohr} discussed the general properties expected for a full (
$nn$, $pp$ and $np$ components)
$T=1$ pairing phonon (appropriate for the region around $^{56}$Ni). Already then, the
evidence (including  both energetics
and transfer data) \cite{nathan} was compelling regarding the major role played by the $T=1$ 
pairing in $N=Z$ nuclei. These ideas were further developed by B\`es and Broglia and culminated
in the review article \cite{bes2}, where a very detailed analysis of isovector pairing vibrations was presented. The role of $T=0$ collective pairing, however, still remained unclear.

A question then arises naturally: is it possible that, although an isoscalar pair condensate
does not seem to form,  $T=0$  collective effects may manifest themselves as a vibrational phonon? 
 With this in mind, we present in this letter a re-analysis of the 
excitation spectrum around the doubly closed-shell $N=Z$ nucleus $^{56}$Ni. Following our 
work on binding energy differences we put an emphasis on the states in odd-odd nuclei, 
and pay particular attention to the subtraction of the full symmetry energy as an integral part
of the average properties that include volume, surface, and Coulomb terms. As we will show, 
the experimental data do not support any marked collectivity in the $T=0$ channel.

\begin{figure}[htbp]
\centerline{\psfig{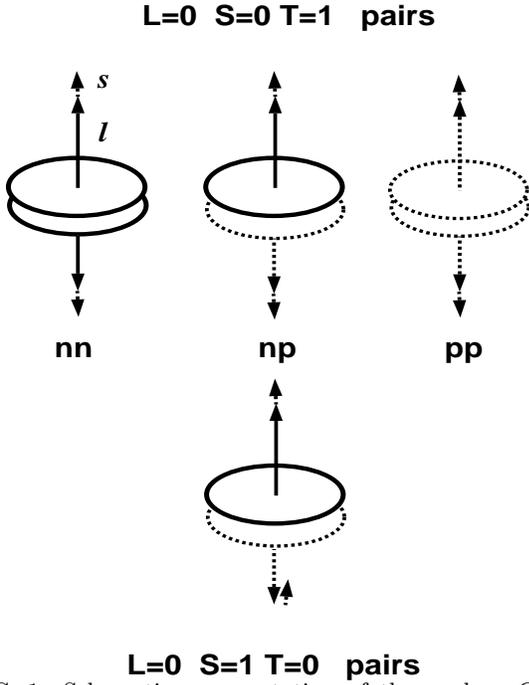}}
\caption{Schematic representation of the nuclear {\sl Cooper} pairs taking part in the pairing correlations discussed in the text }
\label{pairs}
\end{figure}

\begin{figure}[htbp]
\centerline{\psfig{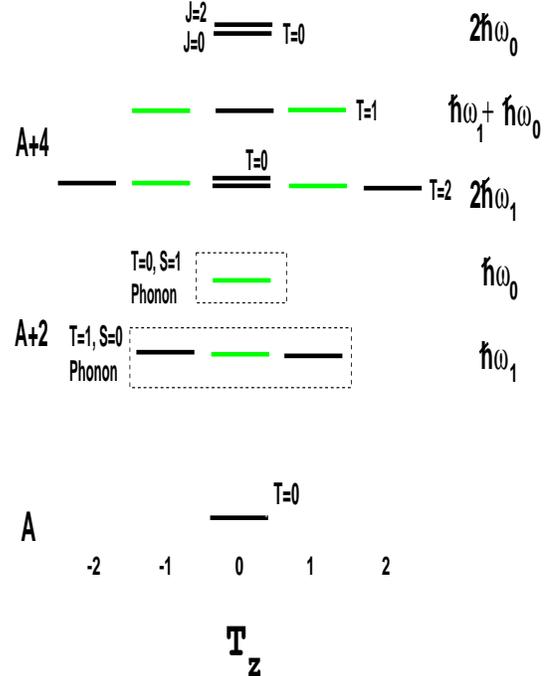}}
\caption{Schematic diagram showing the harmonic spectra for one and two isovector
$(T=1, S=0)$ and isoscalar $(T=0, S=1)$ phonons of energies $\hbar\omega_1$ and
$\hbar\omega_0$ respectively. The example corresponds to the addition
mode on the doubly magic $A(T=0,T_z=0)$ nucleus. Levels in grey are in odd-odd
nuclei.}
\label{phonon}
\end{figure}

Let us start by discussing the nature of the 
excitation spectra arising from competing isovector and isoscalar pairing interactions,
within the framework of two equally degenerate ~ (degeneracy $\Omega$) single-$l$ shells (labelled 1 and 2) separated by an 
energy $\epsilon$ \cite{pang,evans,dussel}.
In this model one considers   $(T=1, L=0, S=0)$ $nn$, $np$, and $pp$ pairs as well as
$(T=0, L=0, S=1)$ $np$ pairs as illustrated in Fig. \ \ref{pairs}. The 
Hamiltonian then involves the scattering of these pairs between the shells

\begin{eqnarray}
H & = & {\epsilon \over 2}( N_2-N_1) \nonumber \\
& & -G_1  \sum_{\mu}(P^{\dag}_{\mu}(1)+ P^{\dag}_{\mu}(2)) 
(P_{\mu}(1)+P_{\mu}(2) )  \nonumber \\
& &-G_0 \sum_{m} (D^{\dag}_{m}(1)+ D^{\dag}_{m}(2))
(D_{m}(1)+D_{m}(2) )
\end{eqnarray}
where $N_{1,2}$ are the numbers of particles in each shell; $P^{\dag}$ refers to the isovector pair creation
operator with isospin projection $\mu$ and
$D^{\dag}$ to isoscalar, ``deuteron-like'', pairs with spin projections $m$.  Here we are interested in the case where the lowest shell is full, representing the doubly magic nucleus, and both pairing strengths ($G_{1,0}$)
are small compared to the single-particle level spacing ($\epsilon$). The collective excitations
are then expected to have a phonon-like spectrum. In the harmonic approximation we have an isovector $(T=1, S=0)$  and an isoscalar 
$(T=0, S=1)$ phonon both of the addition ($A \rightarrow A+2$) and removal ($A \rightarrow A-2$)
types, which are the building blocks to generate the multi-phonon spectra around the doubly magic
nucleus.
The random-phase approximation (RPA) gives for the frequency of these modes \cite{dussel}
\begin{eqnarray}
\hbar\omega_{1,0} & = & \epsilon ( 1 - X_{1,0})^{1/2} 
\end{eqnarray}
with 
\begin{eqnarray}
 X_{1,0}=2G_{1,0}\Omega / \epsilon=G_{1,0}/G_{crit} \nonumber
\end{eqnarray}
where the larger $X$ is, the stronger the correlations. In fact, for $X \sim 1$ the RPA solution
breaks down as the system develops a superconducting phase.
A qualitative spectrum\footnote{It is important to note that this spectrum does not include
volume, surface, and in particular Coulomb and symmetry effects.
These should be considered before a comparison with experiment can be made, and we will do so in the analysis of the data.} corresponding with up to two addition phonons around a closed-shell reference
nucleus is shown in Fig. \ \ref{phonon}. In this example we have assumed that $\hbar\omega_1 < \hbar\omega_0$. The different isospin projections of the isovector addition phonon, $\mu = (-1,0,1)$, correspond to the addition of $(pp,np,nn)$ pairs 
respectively. They give rise to the  isobaric analog multiplet of mass $A+2$,
and angular momentum  $J=0$ at an energy 
$\hbar\omega_1$. In addition, the isoscalar phonon is the $J=1$ state in the odd-odd
$A+2$ nucleus. In going to $A+4$ nuclei, the  two isovector-phonon states give rise to the $T=2$ isobaric multiplet as well as a $T=0$ state in the $N=Z$ even-even nuclei, 
all at an energy of $2\hbar\omega_1$. A $T=1, J=1$ isobaric multiplet at
$\hbar\omega_1 +  \hbar\omega_0$ is obtained by 
adding one phonon of each kind. Finally, the two isoscalar-phonons give
a $T=0, J=0$  and a $T=0, J=2$ state in the $N=Z$,  $A+4$ even-even nucleus at
$2\hbar\omega_0$. 

We now proceed to analyze the experimental data.  We will use $^{56}$Ni as our reference and consider only
even $A$ (even-even or odd-odd) nuclei. The excitation energy of a state of isospin $T$ in a given nucleus $A,Z$  is obtained from
\begin{eqnarray}
E_x(A,T,T_z) & = & BE_{exp} - Surf(A) - Coul(A,T_z) \nonumber \\
& &  -Sym(A,T) - \lambda(A-56) -E_0
\end{eqnarray}
In this equation $BE_{exp}$ includes both the ground-state binding energies from \cite{wapstra}
and, if applicable, the relative excitation energy of the state under consideration
from \cite{toi}.
We use for the surface energy $Surf(A)=-17.3$MeV$A^{2/3}$,
for the Coulomb term $Coul(A)=-0.65$MeV$(A/2-T_z)^2/A^{1/3}$ and the symmetry energy
$Sym(A,T)=-75$MeV$T(T+1) /A$ \cite{aom,jan}.  The volume term enters through the coefficient $\lambda = 15.55$MeV which is
adjusted to give the same frequency for the addition and removal phonons\cite{bohr} and, equivalently,
 sets the
Fermi surface in the middle of the $N=Z=28$ shell gap around $^{56}$Ni.
Finally, $E_0$ is chosen such that $E_x(56,0,0)=0$.
The subtraction of average properties (including an average symmetry term) leaves, in principle, only those effects 
associated with pairing and shell structure\cite{bm2} and can be compared directly with the 
harmonic spectrum discussed above and shown in Fig. \ref{phonon}. In other words, one expects that two-body correlations beyond
the mean-field will survive the subtraction procedure and will generate an excitation spectra whose
properties can be,
at least qualitatively, explained by the Hamiltonian in Eq. (1).

\begin{figure}[htbp]
\centerline{\psfig{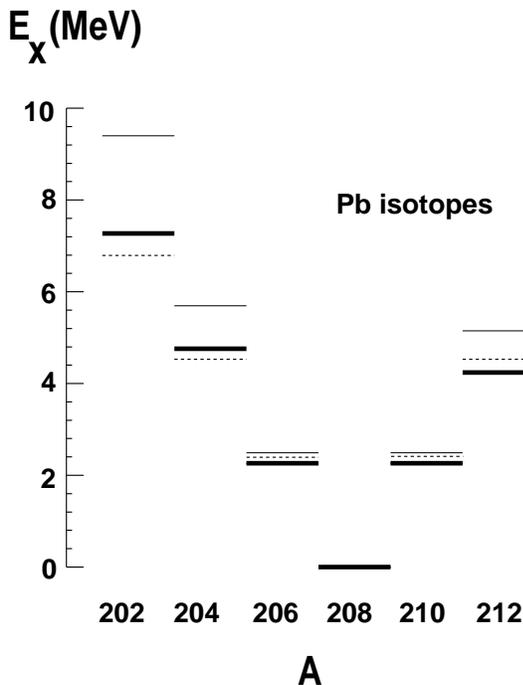}}
\caption{ Experimental excitation spectrum for addition and removal phonons around $^{208}$Pb. 
 The excitation energies of the ground states in even-even Pb isotopes derived from Eq. (3) are shown by the thick lines, those derived 
following [4] by the thin lines and the harmonic approximation by the dashed lines. }
\label{pb208}
\end{figure}

Our method differs slightly from that of Refs. \cite{bes1,bohr}, mainly in the subtraction of
the full symmetry energy term. We thus checked the procedure for nuclei around $^{208}$Pb and obtained for the neutron pairing vibrational phonon
$\hbar\omega \approx 2.3$MeV, in good agreement with the analysis presented in \cite{bm1}.
The results are shown in Fig.  \ref{pb208} where the spectrum obtained following our prescription (thick lines) is compared with that of Ref. \cite{bm1} (thin lines) and the harmonic approximation (dashed lines). It seems that removing the symmetry term produces a spectrum which shows less anharmonicities.

\begin{figure}[htbp]
\centerline{\psfig{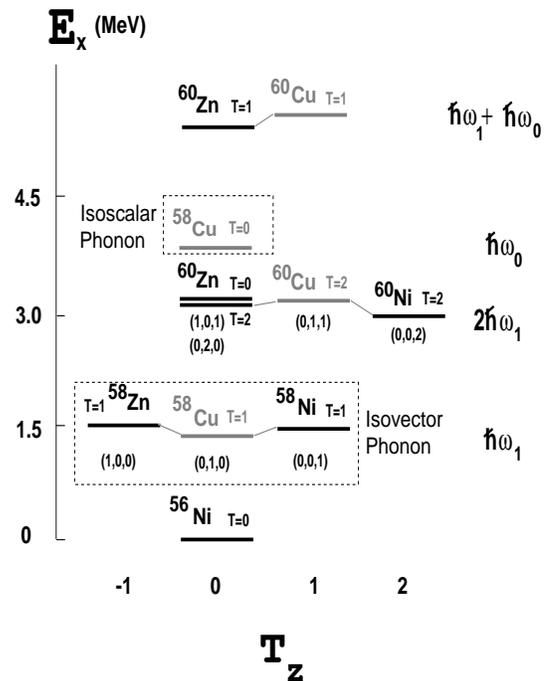}}
\caption{Experimental excitation spectrum for addition phonons around $^{56}$Ni. The excitation energies are derived from Eq. (3). Members of an isobaric
multiplet are joined by thin lines. This figure can be directly compared with
Fig. 2 .}
\label{ni56}
\end{figure}

We now turn our attention to the detailed experimental structure of one and two addition phonons on $^{56}$Ni, shown in Fig. \ref{ni56}   .
In order to make the discussion clearer, the additional labels used in the figure correspond to
 the numbers of $(pp,np,nn)$ pairs. Obviously, $^{56}$Ni is $(0,0,0)$ and, for example,
the ground state of $^{60}$Ni with $T=2$, having two more $nn$ pairs is labelled as $(0,0,2)$.
One immediately realizes the similarity of this spectrum with that of Fig.\ \ref{phonon}.
For the    $T=1$ channel    notice the ``almost'' degenerate energies of the $A=58$ isobaric multiplet, as it 
should be since they correspond to the different isospin projections of the same phonon. The average of those energies gives the frequency
of the $T=1$ phonon, i.e. $\hbar\omega_1 \approx 1.5$MeV.  With a typical single-particle level spacing
energy $\epsilon \approx 4-5 $MeV in this region  we obtain from Eq. 2 a qualitative estimate of
$G_{1}/G_{crit} \approx 0.9$ suggesting a strong collective character of the vibration.

What about the $T= 0$ channel? In this simple picture, the lowest $T=0$ state in 
$^{58}$Cu is obtained from $^{56}$Ni by the addition of the isoscalar $(T= 0, S=1)$ phonon, and therefore
defines the frequency for this mode.  It is obvious by simple inspection of the data that 
this excitation is much higher than for $T=1$. In terms of the phonon picture,  
$\hbar\omega_0 \approx 4$MeV,  which translates into 
$G_{0}/G_{crit} \lesssim 0.2$ and implies, in contrast to the $T=1$ case, very weak (if any) collective
correlations for $T=0$ pairing. In fact, this small lowering of the energy of the $T=0$ state with respect to the single-particle excitation, can be accounted for by a residual $np$ diagonal interaction, $\sim 20$MeV$/A$, rather than any collective effect.

The two-phonon states can now be interpreted in a similar way.
The $T=2$ ground state of
$^{60}$Ni is naturally $(0,0,2)$. In the odd-odd $^{60}$Cu the $(0,1,1)$ component corresponds to an excited
$J=0, T=2$ state (2536 keV) that appears at the expected energy $2\hbar\omega_1 \approx 3$MeV once the average properties (particularly the symmetry energy term)
are subtracted. 
The $T=1$ ground state in $^{60}$Cu is interpreted in this scenario as the $T_z=1$ member of the isobaric multiplet obtained by the coupling of 
a $(T= 0, S=1)$ and a $(T= 1, S=0)$ phonon, at an energy $\hbar\omega_0+\hbar\omega_1 \approx 5.5$MeV. For $N=Z$, the $T=0, J=0$ ground state
in $^{60}$Zn is obtained by a mixture of the $(1,0,1)$ and $(0,2,0)$ components
(resulting in an equal number of $nn$, $np$ and $pp$ pairs).  A $T=2, J=0$ state at an excitation energy 
of 7380 keV from the ground state can be associated with the $T_z=0$ member of the $T=2$ multiplet,
 again, 
once the average properties are subtracted. Likewise, a $T=1$ state at 4913 keV 
(isobaric analog of the $^{60}$Cu ground state) can be interpreted as
the $T_z=0$  member of the $T=1$ multiplet mentioned above.
As discussed earlier, the 
$T=0, J=0$ state obtained by the addition of two isoscalar phonons should lie at
$2\hbar\omega_0 \approx 8$MeV. Its relative excitation energy with respect to the
ground state ($\sim 5$MeV) is so large that it seems unlikely that it
 will mix with the (two $T=1$ phonons) ground state.

A summary of the basic properties of the pairing phonons is given in Table \ \ref{table1},
including the region around $^{40}$Ca where 
a similar behavior is observed.  The pairing strength should
follow a $1/A$ dependence and $G_{crit}=\epsilon/\Omega \sim (1/A^{1/3})/A^{1/3} \sim 1/A^{2/3}$,
therefore $G/G_{crit} \sim 1/A^{1/3}$. It is interesting to note that the data in 
Table \ \ref{table1} seem to follow this mass scaling, thus indicating the same physical mechanism
in the different mass regions.

\begin{table}
\caption{Properties of pairing vibrational phonons. 
\label{table1}}
\begin{center}
\begin{tabular}{ccccc}
Closed Shells & $\hbar\omega $  & $\epsilon $~\tablenotemark[1]  & $G/G_{crit}$& $1/A^{1/3}$~\tablenotemark[2]\\
Nucleus       &    (MeV)        &   (MeV) & \\
\tableline
$^{40}$Ca&  0.8~ $(T=1)$ & 5-6 & $\sim$ 1 & 1 \\
         &  4.9~ $(T=0)$ &     & $\sim$ 0.2\\
$^{56}$Ni&  1.5~ $(T=1)$ & 4-5 & $\sim$ 0.9 & 0.89\\
         &  4.0~ $(T=0)$ &     & $\sim$ 0.2\\
$^{208}$Pb& 2.3 (neutrons) & 3-4& $\sim$ 0.55 & 0.58\\
\end{tabular}
\tablenotetext[1] {from Ref.\ \cite{bm3}.}
\tablenotetext[2] {Normalized to 1 for $^{40}$Ca}
\end{center}
\end{table}

For the purpose of this letter we have only presented the results for two addition phonons.
However, the same picture emerges as one considers more nuclei beyond $^{56}$Ni (and $^{40}$Ca).
The rather soft character of the isovector addition and removal modes, having 
$G_{1} \sim G_{crit}$, strongly suggests the possibility for the multi-phonon system to develop
a permanent deformation in {\sl gauge} space. On the contrary, the isocalar excitations
reflect a weak (if any) collectivity
, $G_{0}/G_{crit} << 1$,  and more likely a
single-particle nature.
Based on these observations, one envisions a coupling scheme whereby the ground
state of any even-even nucleus  is a correlated state of $T=1$ pairs maximally aligned in isospace, i.e. $|A,(T=|T_z|),T_z>$.
In the neighboring odd-odd systems we have: {\it i)} the $|A \pm 2,(T=|T_z|+1),T_z>$ state with one extra  $np(T=1)$
pair and the same pairing correlation properties as the even-even core and  {\it ii)} the
$|A \pm 2,(T=|T_z|),T_z>=   |A,(T=|T_z|),T_z> \otimes ~np(T=0)$ state that is of a single-particle type of coupling.
Due to the predominance of the symmetry term, it is the latter state that is favored as the ground
state of odd-odd nuclei in spite of the extra pairing correlations in the former. Only
in heavier ($A \geq 40$) $N=Z$ odd-odd nuclei, where $2\Delta_{T=1} \sim Sym$, a competition between the two is observed\cite{aom,vogel,Zeldes}.

Finally, it is interesting to consider why for $T=0$  we have $G_0/G_{crit} <<1$ in spite of the fact that the nucleon-nucleon interaction is actually stronger in this channel. Plausible arguments
can be given based on the schematic model discussed above.  Collective effects arise as a competition between the strength $G$, the single-particle energy scale $\epsilon$, and the available phase 
space $\Omega$; this can be expressed in terms of a critical value of the interaction, $\epsilon/2\Omega$.
 Due to the spin-orbit force, the ($T=0, L=0, S=1$) pairs require an extra energy, with respect to the
($T=1, L=0, S=0$) pairs, of the order of the spin-orbit splitting which takes over around mass 20. Therefore, one is left, in the more
appropriate $jj$ coupling scheme, with $J=1$ or $J=j_{max}$ pairs as the most favorable ones. 
In the former, it is possible that 
a combination of both a smaller $G$ due to a reduced spatial overlap, and a smaller degeneracy may hinder the formation of a collective state  . In the latter, it is clear that the phase space is drastically reduced and therefore
 $G_{crit}$ becomes very large.

In conclusion, we have re-visited the pairing vibrations around $^{56}$Ni. Our analysis, built upon
the work of B\`es {\sl et al.},  takes into account the full symmetry energy term and provides a 
natural comparison between the $T=0$ and $T=1$ pairing correlations. The results not only 
confirm the collective behavior of the isovector pairing vibrations but indicate a single-particle character for the isoscalar channel.

We would like to thank Prof. Daniel B\`es for his comments on the manuscript.
This work was supported by the Division of Nuclear Physics of the
US Department of Energy, under contract DE-AC03-76SF00098.



\begin{references}
\bibitem{aom} A.O.Macchiavelli et al. submitted to Phys. Rev. C, and Los Alamos Preprint nucl-th/9907121.
\bibitem{vogel} P.Vogel, Nucl. Phys. in press,  and Los Alamos Preprint nucl-th/9805015.
\bibitem{bes1} D.R.B\`es and R.A.Broglia, Nucl. Phys. {\bf 80}, 289 (1966).
\bibitem{bm1} A.Bohr and B.R.Mottelson, Nuclear Structure, 
Benjamin-New York, (1969 and 1975), Vol II, pag. 645.
\bibitem{bohr} A.Bohr, Proceedings of Dubna Symposium on Nuclear Structure, IAEA (1969), pag. 179.
\bibitem{nathan}O.Nathan,  Proceedings of Dubna Symposium on Nuclear Structure, IAEA (1969), pag. 191.
\bibitem{bes2}D.R.B\`es et al. Phys. Rep. {\bf 34C}, 1 (1977), and references therein.
\bibitem{pang} S. Pang, Nucl. Phys.  {\bf A128}, 497 (1969).   
\bibitem{evans} J. Evans et al., {\bf A367}, 77 (1981). 
\bibitem{dussel} G. Dussel et al.,  Nucl. Phys. {\bf A153}, 469 (1970)
\bibitem{wapstra} G.Audi and A.H.Wapstra,  Nucl. Phys. {\bf A565}, 1 (1993)
\bibitem{toi}Table of Isotopes 8th Edition, R.B.Firestone and V.S.Shirley Eds.,John Wiley \& Sons Inc. (1996), Vol. I.
\bibitem{jan}J.Janecke, Nucl. Phys. {\bf A73}, 97 (1965).
\bibitem{bm2} A.Bohr and B.R.Mottelson, {\it op cit} Vol I , pag. 145.
\bibitem{Zeldes} N.Zeldes and S.Liran,  Phys. Lett. {\bf B62}, 12 (1976)
\bibitem{bm3} A.Bohr and B.R.Mottelson , {\it op cit} Vol I , pags. 325, 328.

\end{references}
\end{document}